\documentstyle[aps]{revtex}
\begin{document}
\title{Chern-Simons Gravity: From 2+1 to 2n+1Dimensions \thanks{%
Lecture presented at the {\em XX Encontro de Fisica de Part\'{i}culas e
Campos}, Sao Lourenco, Brazil, October 1999, and at the {\em Fifth La
Hechicera School,} Merida, Venezuela, November 1999.   }}
\author{Jorge Zanelli$^{1,2}$}
\address{$^{1}$Centro de Estudios Cient\'{\i }ficos (CECS), Casilla 1469,
Valdivia,
Chile \\
$^{2}$Departamento de F\'{i}sica, Universidad de Santiago de Chile, Casilla
307, Santiago 2, Chile}
\date{}
\maketitle

\begin{abstract}
These lectures provide an elementary introduction to Chern Simons Gravity
and Supergravity in $d=2n+1$ dimensions.
\end{abstract}

\section{Introduction}

The present situation of High Energy Physics is both exciting and
paradoxical. On one side we have a monumental theoretical machinery built on
the most beautiful ideas --strings, membranes, dualities, $M-$theory,
AdS-CFT correspondence--, with hardly any experimentally testable
predictions for the near future. On the other hand, there is a growing
corpus of puzzling observations coming mostly from astrophysics, including
gamma ray bursts, missing mass, microwave inhomogeneities, indications of an
accelerated expansion, which defy present standard wisdom about the
universe, acquired through observations of our more immediate neighborhood.
The new observations are for the most part surprising, and one could expect
by now that the surprises will continue. In an effort to cope with the new
data, a number of exotic proposals have been put forward. Thus, in recent
times notions such as {\em quintessence} and other nonminimal extensions of
General Relativity ({\bf GR}) including a nonvanishing cosmological
constant, and extra dimensions at macroscopic scales have gained acceptance
in respectable journals.

In this situation, a reasonable attitude for a theoretical physicist would
be to critically examine all consistent alternatives to our time-honored
foundations. In this vein, here we discuss a class of gravity theories which
share the essential geometric foundation of GR:

\begin{itemize}
\item  {\bf a}) General covariance

\item  {\bf b}) Second order field equations for the metric

At the same time, other features of GR are relaxed:

\item  {\bf c}) Spacetime is allowed to have any number of dimensions

\item  {\bf d}) Spacetime is not necessarily asymptotically flat

\item  {\bf e}) The action is extremized under independent variations of the
metric and affine connection
\end{itemize}

Condition (c) is absolutely necessary in the context of string theory and
(d) is most natural in the light of the recently discovered AdS-CFT
correspondence.

Requirement (e), on the other hand, allows for independent propagation of
the metric and affine structures of spacetime. This is satisfactory in view
of the fact that the metric and affine features of spacetime are
geometrically independent: one has to do with the measurement of distances
while the other relates to parallel transport. A natural way to implement
condition (c) is to use the first order formalism, where the independent
fields are the vielbein one-form $e^{a}=e_{\mu }^{a}dx^{\mu },$and the spin
connection $\omega ^{ab}=\omega _{\mu }^{ab}dx^{\mu }$.

\medskip The family of gravity theories obtained with these postulates where
studied by Lovelock in the early 70's \cite{Lovelock}. The five dimensional
case had already been discussed by Lanczos in the late 30's \cite{Lanczos}.
More recently, Zwiebach \cite{Zwiebach} and Zumino \cite{Zumino} showed the
Lanczos-Lovelock {\bf (LL)} theories to be appropriate to describe the
effective low-energy gravity theory found in the weak coupling limit of
string theory.

In their simplest version, the {\bf LL} theories have the same fields,
symmetries and local degrees of freedom as ordinary gravity. The action is a
polynomial of degree $[d/2]$ in curvature\footnote{%
Here $[x]$ is the integer part of $x$.}, which can also be written in terms
of the Riemann curvature $R^{ab}=d\omega ^{ab}+\omega _{c}^{a}\omega ^{cb}$
and the vielbein as\footnote{%
Wedge product between forms is understood throughout.}

\begin{equation}
I_{G}=\kappa \int \sum_{p=0}^{[d/2]}\alpha _{p}L^{(p)},  \label{Lovaction}
\end{equation}
where $\alpha _{p}$ are arbitrary constants, and $L^{(p)}$ is given by

\begin{equation}
L^{(p)}=\epsilon _{a_{1}\cdots a_{d}}R^{a_{1}a_{2}}\!\cdot \!\cdot \!\cdot
\!R^{a_{2p-1}a_{2p}}e^{a_{2p+1}}\!\cdot \!\cdot \!\cdot \!e^{a_{d}}.
\label{Lovlag}
\end{equation}

What makes the LL theories so special is the fact that they comply with the
requirement b) above. An arbitrary Lagrangian density constructed with the
metric and curvature tensors, on the contrary, would give rise to fourth
order field equations for the metric, in general. A powerful reason to
choose the Einstein-Hilbert ({\bf EH})action in four dimensions is that the
Einstein equations are second order. This feature of the EH action~stems
from the fact that the EH Lagrangian in {\em two dimensions} is the density
of a topological invariant: the Euler characteristic of the manifold.
Similarly, the LL theories in $d$-dimensions are linear combinations (with
arbitrary coefficients) of the Euler densities of all dimensions below $d$.
Thus, General Relativity is a particular case of LL theory.

\subsection{Drawbacks of the Generic LL Action}

In spite of their nice features, the LL theories suffer from an original
sin: they are endowed with a collection of indeterminate dimensionful
parameters $\alpha _{p}$, $p=1,...,\left[ \frac{d}{2}\right] $. This has two
puzzling consequences already at the classical level:

\begin{itemize}
\item  ({\bf i}) The theories have a large number of physical parameters
which should be experimentally determined. This would make gravity less
interesting as a fundamental theory, because it would have more natural
constants, like $G_{Newton\text{ }}$and the cosmological constant $\Lambda $%
, to be adjusted.

\item  ({\bf ii}) The field equations admit solutions which have
indeterminate spacelike dependence and timelike evolution. This results from
the fact that the field equations are polynomials of degree $p$ in the
derivatives of the metric $(\partial g)$. Thus, the values of the velocity
can jump between different roots of the equation arbitrarily and still
extremize the action\cite{TZ}, \cite{HTZ}.
\end{itemize}

A consequence of (ii) is the fact that the Legendre transformation from the
Lagrangian to the Hamiltonian cannot be performed in general, making the
canonical quantization program ill-defined. Another problem related with the
second issue is the existence of several vacua with different topologies.
This could be an interesting novel feature of these theories, were it not
for the fact that the perturbation expansions around the different vacua
give rise to completely different theories and typically contain ghosts \cite
{BD}.

There is one more reason to find the presence of the large number of
coefficients in the Lagrangian undesirable. The bare $\alpha _{p}$'s are
dimensionful constants and therefore could receive quantum
corrections~beyond control, making the possibility of constructing a quantum
theory of gravity even more remote than in standard GR. This would be so
unless the values of these constants are protected by some fundamental
symmetry, like the zero mass for the photon, or the equal number of quarks
and leptons, which are ``protected'' by gauge invariance.

Thus, it would be interesting to find a ``natural'' way to fix the $\alpha $%
's. Moreover, if the criterion that fixes these coefficients is based on
some symmetry principle, and possibly protect them from renormalization with
a reasonable symmetry principle. In this scenario, one finds two special
families, which stand out among all LL theories. They correspond to a
Born-Infeld -like theory in even dimensions, and the Chern-Simons theory for
the anti-de Sitter ({\bf AdS}) gauge group for odd D. In section {\bf III}
we review these theories in general. In section {\bf IV} the CS case and
their supersymmetric extensions are analyzed in greater depth. The next
section is just a cursory review of nonabelian CS theories which can be
skipped by the experts and by those eager to get to the juicy stuff.

\section{Chern-Simons Theory in 3 Dimensions}

Chern-Simons ({\bf CS}\footnote{There is a vast literature on this subject. 
For an extensive treatment of CS theories, see e.g., \cite{Nakahara}, or 
\cite{EGH}.}) theory has a curious history. It was discovered in
the context of anomalies in the 70's and used as a rather exotic toy model
for gauge systems in 2+1 dimensions ever since \cite{DJT} . It was only by
the mid 80's that it was realized that ordinary Einstein gravity in 2+1
dimensions is a natural example of a {\bf CS} system, especially through the
work of Witten \cite{Witten}. As it turns out, {\bf CS} systems are more
conspicuous than it might seem at first sight. General Relativity in 2+1
dimensions (with or without cosmological constant) is a {\bf CS} system (for
ISO(2,1) or SO(2,2) groups, respectively); any ordinary mechanical system in
Hamiltonian form can be viewed as an abelian {\bf CS} system in 0+1
dimensions\cite{STZ}. This way of looking at mechanical systems is not
completely absurd and it even sheds some light into ancient problems such as
the justification for the old quantization rule of Bohr and Sommerfeld.

In retrospect, we can see that the key to the construction of the {\bf CS}
form (in three dimensions) is the following: the Pontryagin form

\begin{equation}
P=\text{Tr}[{\bf F\wedge F}],  \label{Pontryagin}
\end{equation}
is {\bf closed}

\begin{equation}
dP=0.  \label{dP=0}
\end{equation}
By Poincar\'{e}'s lemma, $P$ is {\bf locally exact}, that is, it is always
possible to write it in an open neighborhood as the exterior derivative of a
3-form

\begin{equation}
P=dL.  \label{P=dL}
\end{equation}
Thus, the 3-form~$L$ is the {\bf Chern-Simons} Lagrangian. Clearly, this
idea can be generalized to higher (odd) dimensions, and for other integral
topological invariants, like the Euler characteristic. This is precisely
where the connection with the LL theory can be found: if one asks, {\em what
is the 3-form whose exterior derivative is the 4-dimensional Euler density?}%
, the answer is the Einstein-Hilbert action with nonzero cosmological
constant.\medskip We now briefly review a few facts about standard
3-dimensional {\bf CS} systems.

The idea is to find a three-form $L_{CS}$ such that

\begin{equation}
dL_{CS}=\text{Tr}[{\bf F\wedge F}],  \label{dL}
\end{equation}
where 
\begin{equation}
{\bf F}={\bf dA+A\wedge A}  \label{F}
\end{equation}
is the curvature (field strength) in the adjoint representation and ${\bf A}$
is a Lie algebra-valued connection1-form.

Let ${\bf G}$ be the gauge group and ${\em G}$ its Lie algebra generated by
the matrices ${\bf T}_{a}$, such that $[{\bf T}_{a},{\bf T}_{b}]=C_{ab}^{c}%
{\bf T}_{c}$. Under the action of the gauge group, the connection

\[
{\bf A=}A_{\mu }^{a}{\bf T}_{a}dx^{\mu } 
\]
transforms as 
\begin{equation}
{\bf A\rightarrow A}^{\prime }=g^{-1}{\bf A}g+g^{-1}dg,  \label{gaugeA}
\end{equation}
where the 0-form $g(x)$ is an element of $G$. Then, the curvature changes as 
\begin{equation}
{\bf F\rightarrow F}^{\prime }=g^{-1}{\bf F}g,  \label{gaugeF}
\end{equation}
and it is easily shown, using the cyclic property of the trace, that Tr$[%
{\bf F\wedge F}]$ is invariant under (\ref{gaugeA}, \ref{gaugeF}). From (\ref
{dL}), the {\bf CS} Lagrangian is found to be\medskip

\begin{equation}
L_{CS}=\text{Tr}[{\bf A}\wedge d{\bf A}+\frac{2}{3}{\bf A}\wedge {\bf A}%
\wedge {\bf A}].  \label{CS3N-A}
\end{equation}

\subsection{Gauge Invariance}

It is easily checked that the{\bf \ CS} action is invariant under gauge
transformations. First observe that under a gauge transformation of the form
(\ref{gaugeA}), the right hand side of (\ref{dL}) does not change and
therefore $\delta (dL_{CS})=0$%
\[
d\delta L_{CS}=0. 
\]

In other words, under a variation of the fields that leaves the Pontryagin
form invariant, the {\bf CS} Lagrangian changes by a closed form. Thus,
provided the change $\delta L_{CS}$ approaches zero sufficiently fast at the
spacetime boundary, the {\bf CS} action should be invariant as well.
Substituting (\ref{gaugeA}) in (\ref{CS3N-A}) one finds 
\begin{equation}
L_{CS}({\bf A}^{\prime })=L_{CS}({\bf A})-d\text{Tr}[dgg^{-1}{\bf A}]-\frac{1%
}{3}\text{Tr}[(g^{-1}dg)^{3}].  \label{L(A')}
\end{equation}
Then the action changes as 
\begin{equation}
I_{CS}[{\bf A}^{\prime }]=I_{CS}[{\bf A}]-\int_{\partial M}\text{Tr}[dgg^{-1}%
{\bf A}]-\frac{1}{3}\int_{M}\text{Tr}[(gdg^{-1})^{3}].  \label{deltaI}
\end{equation}
This raises an important issue: is the action{\em \ really} invariant under
the gauge transformation (\ref{gaugeA})? The answer seems to be no, unless $%
g\rightarrow 1$ sufficiently fast to cancel the second term, and some other
miracle makes the third term also vanish. The first condition one can always
demand because it is part of the rules of the game in any variational
problem that the fields satisfy appropriate boundary conditions, and that
necessarily restricts the type of field transformations allowed at the
spacetime boundary.

The last term in (\ref{L(A')}) is closed and therefore, provided there are
no topological contrivances, this term can be expressed locally as the
exterior derivative of some 2-form which depends on $g(x)$. It is obvious
that this term cannot be made to vanish simply by imposing some asymptotic
condition on the gauge transformation $g$. In fact, there are some
interesting simple examples in which the integral of this last term doesn't
vanish. For instance, if the manifold $M$ is topologically a 3-sphere and
the gauge group is $SU(2)$, the last term in (\ref{deltaI}) is $4\pi ^{2}N$,
where $N$ is the {\bf winding number} of the mapping $g:M\rightarrow SU(2)$.

The transformation law (\ref{deltaI}) tells us that the action changes by a
surface term and possibly by a functional of $g(x)$. Although none of these
terms can alter the field equations, they can change the global properties
of the theory, like the definition of conserved charges. They also provide
different weight for topologically different configurations in the quantum
theory as defined by the path integral. At any rate, the action $I_{CS}[{\bf %
A}]$ is genuinely gauge invariant if the manifold has no boundary ($\partial
M=0$), or the gauge transformation goes to zero at the boundary fast enough,
and the topology of the mapping $g:M\rightarrow {\bf G}$ is trivial. These
conditions are met by gauge transformations which are everywhere
infinitesimally close to the identity, e.g., $g=1+\lambda ^{a}(x){\bf T}_{a}$%
, with $\lambda ^{a}(x)<<1$. This is sufficient for most practical purposes
in the study of {\bf CS} systems as field theories.

\subsection{Field equations}

Now that we have a good action principle for the connection ${\bf A}$, it is
natural to ask, what does it describe? One way to answer this is to study
the field equations. Varying the action yields 
\[
\delta I_{CS}[{\bf A}]=2\int_{M}\text{Tr}[{\bf T}_{a}{\bf T}_{b}]F^{a}\wedge
\delta A^{b}{\bf \ -}\int_{\partial M}\text{Tr}[{\bf T}_{a}{\bf T}%
_{b}]A^{a}\wedge \delta A^{b}. 
\]

Here we see that the condition of having an extreme under arbitrary
variations $\delta A^{b}$implies 
\begin{equation}
F^{a}=0,  \label{Fo}
\end{equation}
provided the group algebra is such that 
\[
\gamma _{ab}\equiv \text{Tr}[{\bf T}_{a}{\bf T}_{b}] 
\]
is a non-singular matrix, which is always the case for a {\bf semisimple}
Lie algebra in the adjoint representation and $\gamma _{ab}$ is the Killing
metric.

Semisimple algebras are those which do not contain invariant abelian
subalgebras (roughly, those that {\em cannot} be written as ${\cal G}$ $=%
{\cal G}_{0}\oplus {\cal U}$, where ${\cal U}$ is abelian). Semisimple
algebras correspond to the classical groups $SO(n)$, $SU(n)$, $Sp(2n)$, and
other more exotic choices such as $OSp(n,m)$, $USp(p,q)$, $E_{8}$, etc.
There is an important exceptional Lie algebra which {\bf is not} semisimple
and yet there is a faithful representation for which $\gamma _{ab}$ is
nondegenerate. This is the case of the Poincar\'{e} group in 2+1 dimensions, 
$ISO(2,1)$, whose algebra is $so(2,1)\oplus R^{2,1}$, where $R^{2,1}$ is the
group of translations in 2+1 dimensions. This exception to the rule allows
writing the Einstein-Hilbert Lagrangian as a {\bf CS} 3-form for the
Poincar\'{e} group, and is the key to the quantizability of gravity in 2+1
dimensions \cite{Witten}.

The field equations (\ref{Fo}) look strikingly simple and not much could be
expected from them. In fact, there is a well known result according to which
if on a homotopically trivial open set $B$, 
\[
d{\bf A}+{\bf A}\wedge {\bf A}=0, 
\]
then ${\bf A}$ can be written as a gauge transformation of a trivial
connection, 
\begin{equation}
{\bf A}=g^{-1}dg\text{ everywhere in }B.  \label{pure gauge}
\end{equation}

In other words, by a gauge transformation it is always possible to take $%
{\bf A}\rightarrow 0$ on any small patch of $M$. Thus, the field
configurations described by the equation (\ref{Fo}) would be trivial unless
there are topological obstructions which prevent (\ref{pure gauge}) from
being valid globally throughout the entire manifold $M$ even if it remains
valid on any small open set $B$. This is in fact what happens with gravity
in 2+1 dimensions, where the field equations are of the form (\ref{Fo}) and
still one can have nontrivial solutions such as black holes and
gravitational collapse. At any rate, what remains true is the fact that a 
{\bf CS} system in 2+1 dimensions has no local degrees of freedom that could
propagate. In higher dimensions, however, {\bf CS }systems possess
propagating local degrees of freedom and the situation is similar to that of
a standard gauge theory.

\subsection{Generalization to Higher Dimensions}

In spite of their interesting mathematical structure, these
three-dimensional theories might seem too unrealistic as models for our
world. We are now going to see how these ideas are extended to higher
dimensions. The essential ingredient in the construction of CS theories in
higher dimensions is the existence of a $2n-$ form 
\begin{equation}
Q_{2n}({\bf A})=\gamma _{a_{1}...a_{n}}F^{a_{1}}\wedge F^{a_{2}}\wedge \cdot
\cdot \cdot \wedge F^{a_{n}},  \label{Fn}
\end{equation}
which is closed, 
\[
dQ_{2n}=0\text{ ,} 
\]
and invariant under a gauge transformation ${\bf A}\rightarrow {\bf A}%
^{\prime }=g^{-1}{\bf A}g+g^{-1}dg$, 
\[
Q_{2n}({\bf A}^{\prime })=Q_{2n}({\bf A}). 
\]

It is straightforward to show that the invariants of the form 
\[
Q_{2n}({\bf A})=\underbrace{\left\langle {\bf F}\wedge {\bf F}\wedge \cdot
\cdot \cdot \wedge {\bf F}\right\rangle }_{\mbox{n-times}} 
\]
\medskip satisfy these requirements, where we have defined 
\[
\gamma _{a_{1}...a_{n}}\equiv \left\langle T_{a_{1}}T_{a_{2}}\cdot \cdot
\cdot T_{a_{n}}\right\rangle , 
\]
and $\left\langle ...\right\rangle $ stands for a trace operation in an
appropriate representation of the Lie algebra {\em G}. ~The invariants of
the form (\ref{Fn}) are in one-to-one correspondence with the $n$th rank
invariant tensors $\gamma _{a_{1}...a_{n}}$ which could be constructed for a
given gauge group. The number of such tensors is rather small in general and
is related to the number of Casimir invariants of the group. (In the next
two sections we will discuss specific realizations of these brackets, so
most of the mystery will be dissipated shortly.) From now on, we will not
explicitly write the wedge symbol unless there is an ambiguity, so we will
also write $Q_{2n}$ as 
\begin{equation}
Q_{2n}({\bf A})=\left\langle {\bf F}^{n}\right\rangle .  \label{Q2n}
\end{equation}

These invariants belong to the family of {\bf characteristic classes }known
as the Chern-Weil invariants. These classes themselves define integral
invariants related to the topological properties of the maps that can be
established between manifold $M$ and the group ${\bf G}$.

The equation analogous to (\ref{dL}) now reads 
\begin{equation}
dL_{CS}^{2n-1}=\left\langle {\bf F}^{n}\right\rangle ,
\end{equation}
and its solution can be written as 
\begin{equation}
L_{CS}^{2n-1}=\frac{1}{(n+1)!}\int_{0}^{1}dt\left\langle {\bf A}(td{\bf A}%
+t^{2}{\bf A}^{2})^{n-1}\right\rangle +\alpha ,  \label{L(2n-1)}
\end{equation}
where $\alpha $ is an arbitrary closed ($2n-1$)-form ($d\alpha =0$).

Under a gauge transformation of the form (\ref{gaugeA}), the Chern-Simons
form (\ref{L(2n-1)}) changes as 
\begin{equation}
L_{CS}^{2n-1}({\bf A}^{\prime })=L_{CS}^{2n-1}({\bf A})+d\beta +(-1)^{n-1}%
\frac{n!(n-1)!}{(2n-1)!}\left\langle (g^{-1}dg)^{2n-1}\right\rangle ,
\label{DeltaL(2n-1)}
\end{equation}
where the ($2n-1$)-form $\beta $ is a function of ${\bf A}$ and depends on $%
g $ through the combination $g^{-1}dg$. Thus, the action

\[
I_{CS}^{2n-1}[{\bf A}]=\int_{M}L_{CS}^{2n-1}, 
\]
describes a gauge theory for the group ${\bf G}$, which under a finite gauge
transformation changes as the integral of (\ref{DeltaL(2n-1)}). The second
term in the RHS of (\ref{DeltaL(2n-1)}) gives rise to a boundary term, and
the third is proportional to the winding number. Again, one can see that
under an infinitesimal gauge transformation (connected to the identity) of
the form 
\[
\delta {\bf A}=-\nabla \lambda ,\text{ where }\lambda <<1, 
\]
the action changes by a surface term, which can be set to zero under the
appropriate boundary conditions.

\section{\protect\medskip General Relativity as a Chern-Simons Theory}

So far, the best description of our universe at large scale is general
relativity with the Einstein-Hilbert {\bf (EH)} action defined on a
four-dimensional spacetime 
\begin{equation}
I=\int_{M_{4}}\sqrt{-g}(R-2\Lambda )d^{4}x,  \label{EH}
\end{equation}
where $R$ is the Ricci scalar curvature and $\Lambda $ is the cosmological
constant. More than sixty years of ~frustrated efforts to quantize this
theory can explain the immediate attention drawn by Witten's classical
observation that gravity in 2+1 spacetime dimensions is an exactly solvable
model\cite{Witten}. This result means that the quantum theory can be
completely and explicitly spelled out, that is, all correlation functions
or, alternatively, its entire Hilbert space, can be known. This is
remarkable since in just one more spatial dimension, the quantization
problem becomes intractable. It could be argued that quantization of 2+1
gravity is no big deal since the theory has no propagating degrees of
freedom and therefore its quantum description is like that of a system of
point particles. Although this is certainly an important simplification, the
key to the proof of solvability is the fact that 2+1 gravity is a{\bf \ CS}
system and hence, it has all the nice features of a gauge theory. Gravity in
3+1 dimensions, on the other hand cannot be construed as a gauge system of
the Poincar\'{e} or (A-)dS groups, and this is a serious limitation for it
quantization. In what follows we will cast 2+1 gravity as a CS system and
will see how this can be generalized to higher dimensions.

The three-dimensional EH action analogous to (\ref{EH}) can be cast as a
first order theory, in which only first order derivatives occur, by using
form language, 
\begin{equation}
I[\omega ,e]=\frac{1}{2}\int \epsilon _{abcd}\left( R^{ab}-\frac{1}{6}%
\Lambda e^{a}e^{b}\right) e^{c}e^{d}.  \label{EH'}
\end{equation}

Here $R^{ab}$ is the curvature two-form $R_{\text{ }b}^{a}=d\omega _{\text{ }%
b}^{a}+\omega _{\text{ }c}^{a}\omega _{\text{ }b}^{c}$ (here the wedge
product symbol ($\wedge $) has been suppressed), and the dynamical fields
are the vielbein $e^{a}=$ $e_{\mu }^{a}(x)dx^{\mu }$, and the Lorentz (spin)
connection $\omega _{\mu }^{ab}=\omega _{\mu }^{ab}(x)dx^{\mu }$. Note that
the action (\ref{EH'}) depends on the fields and their {\em first
derivatives only} (this is a consequence of the fact that only exterior
derivatives are used throughout and therefore higher order derivatives with
respect to one coordinate can never occur). When this action is varied with
respect to $e^{d}$ and $\omega ^{ab}$, the following field equations are
found 
\begin{eqnarray}
\epsilon _{abcd}\left( R^{ab}-\frac{1}{3}\Lambda e^{a}e^{b}\right) e^{c} &=&0
\label{Einstein} \\
\epsilon _{abcd}T^{c}e^{d} &=&0.  \label{Torsion}
\end{eqnarray}
The first expression are the usual Einstein's equation, which relates the
curvature to the metric, while the second implies vanishing torsion, $%
T^{a}:=De^{a}=$ $de^{a}+\omega _{\text{ }b}^{a}e^{b}=0$, which is usually
postulated in general relativity. Solving this second equation for $\omega
^{ab}$ as a function of $e$, $de$ and $e^{-1}$ and substituting $\omega
^{ab}(e)$ into (\ref{Einstein}) the standard second order form of the
Einstein equations is obtained. The action (\ref{EH'}) is the first order
formulation of the EH theory and it is the most general local $4$-form
invariant under Lorentz rotations in the tangent space, constructed out of
the vielbein, the spin connection and their exterior derivatives only \cite
{Regge,Zumino}. This suggests that in other dimensions, the same
prescription could be applied to construct the action for the gravitational
field.

\subsection{Beyond the Einstein-Hilbert action}

In order to describe the gravitational field for $d>4$ one could assume the
spacetime geometry as given by the same {\bf EH} action (\ref{EH}) --with or
without cosmological constant-- integrated now over $d$ dimensions. One of
the reasons for the universal appeal of the EH action is that it yields
second order field equations, but as we already mentioned, other attractive
alternatives exist in higher dimensions of the form (\ref{Lovaction}){\bf .}

The Lanczos-Lovelock ({\bf LL}) Lagrangian is the most general local $d$%
-form invariant under Lorentz rotations of the tangent space, constructed
out of the vielbein, the spin connection and their exterior derivatives
without using other structures \cite{Regge}. In particular, the metric, the
inverse vielbein or the Hodge-* dual\footnote{%
The Hodge $*$-operation relates a $p$-form and a $(d-p)$-form through its
action on the basis 
\begin{eqnarray*}
\ast (dx^{\mu _{1}}\cdot \cdot \cdot dx^{\mu _{p}}) &=&\frac{1}{(d-p)!}\sqrt{%
|g|}g^{\mu _{1}\nu _{1}}\cdot \cdot \cdot g^{\mu _{p}\nu _{p}}\times \\
&&\epsilon _{\nu _{1}\cdot \text{ }\cdot \text{ }\cdot \nu _{d}}dx^{\nu
_{p+1}}\cdot \cdot \cdot dx^{\mu _{d}}.
\end{eqnarray*}
Note that this expression involves explicitly the inverse of the metric and
its inverse.} are never used in the construction, ensuring that only first
order field equations for $e$ and $\omega $ can be produced. If the
torsion-free condition is assumed, the equations for the metric become
second order.

\subsection{Gravity as an (anti-)de Sitter CS Theory}

The coefficients in the {\bf LL} action can be made dimensionless by the
redefinition $\alpha _{p}\rightarrow \alpha _{p}l^{d-2p}$, where $l$ is a
parameter with dimensions of length. Then, the {\bf LL} Lagrangian reads 
\begin{equation}
L=\sum\limits_{p=0}^{[d/2]}\alpha _{p}l^{2p-d}\epsilon
_{a_{1}...a_{d}}R^{a_{1}a_{2}}\cdot \cdot \cdot
R^{a_{2p-1}a_{2p}}e^{a_{2p+1}}\cdot \cdot \cdot e^{a_{d}}.  \label{Lovelock}
\end{equation}
The embarrassing freedom to choose the $\alpha _{p}$'s arbitrarily can be
drastically cut by the following observation: The field equations for $e$
and $\omega $ involve $R^{ab}$, $T^{a}$ and $e^{a}$. Taking the covariant
exterior derivatives of these equations and using the Bianchi identities,
new algebraic relations between the curvature and torsion tensors are
produced, which would generically introduce nonholonomic restrictions on the
degrees of freedom of the theory . This is so for all choices of $\alpha
_{p} $'s, except in two cases, for which no new constraints on the geometry
are generated \cite{HDG}:

\begin{itemize}
\item  $d=2n$: The Born-Infeld {\bf (BI)} case, 
\begin{equation}
\alpha _{p}=\kappa \left( 
\begin{array}{l}
n \\ 
p
\end{array}
\right) ,o\leq p\leq n  \label{alpha2n}
\end{equation}

\item  $d=2n-1$: The AdS Chern-Simons {\bf (AdS-CS)} case, 
\begin{equation}
\alpha _{p}=\frac{\kappa }{d-2p}\left( 
\begin{array}{l}
n-1 \\ 
p
\end{array}
\right) ,0\leq p\leq n-1.  \label{alpha2n-1}
\end{equation}
\end{itemize}

The reference to BI in the first case stems from the fact that, in that
case, the Lagrangian reads 
\begin{eqnarray}
L &=&\kappa \cdot \epsilon _{a_{1}...a_{d}}\left( R^{a_{1}a_{2}}+\frac{1}{%
l^{2}}e^{a_{1}}e^{a_{2}}\right) \cdot \cdot \cdot \left( R^{a_{d-1}a_{d}}+%
\frac{1}{l^{2}}e^{a_{d+1}}e^{a_{d}}\right) ,  \nonumber \\
&=&\kappa \cdot pfaff\left[ R^{ab}+\frac{1}{l^{2}}e^{a}e^{b}\right] ,
\label{BI} \\
&=&\kappa \cdot \sqrt{\det \left[ R^{ab}+\frac{1}{l^{2}}e^{a}e^{b}\right] },
\nonumber
\end{eqnarray}
which is reminiscent of Lagrangian for the Born-Infeld electrodynamics

The odd-dimensional case is referred to as AdS-CS because that Lagrangian is
of the CS family and it can be cast in a form which is manifestly invariant
under local AdS transformations. This can be seen as follows. Consider the
array of 1-forms 
\begin{equation}
W^{AB}=\left[ 
\begin{array}{ll}
\omega ^{ab} & l^{-1}e^{a} \\ 
-l^{-1}e^{b} & 0
\end{array}
\right] ,  \label{W}
\end{equation}
where the indices $a,b,...=1,2,..d$ and $A,B,...=1,2,..d+1$. It is a simple
exercise to show that this connection defines a curvature 2-form given by 
\begin{eqnarray}
\overline{R}^{AB} &=&dW^{AB}+W_{\text{ }C}^{A}W^{CB}  \nonumber \\
&=&\left[ 
\begin{array}{ll}
R^{ab}+l^{-2}e^{a}e^{b} & l^{-1}T^{a} \\ 
-l^{-1}T^{b} & 0
\end{array}
\right] ,  \label{Rbar}
\end{eqnarray}
where the $A,B,...$ indices are raised and lowered using the AdS metric%
\footnote{%
All that it is said in this lecture about AdS can be easily carried over to
the de Sitter case. It is necessary to change a few signs at the right
places.}, 
\begin{equation}
\eta _{AB}=\text{diag}(-1,-1,+1,....,+1).  \label{AdSeta}
\end{equation}
The AdS curvature $\overline{R}^{AB}$ can be used to construct the {\bf \
Euler form} in $2n$ dimensions as 
\begin{equation}
{\cal E}_{2n}=\epsilon _{A_{1}...A_{2n}}\overline{R}^{A_{1}A_{2}}\cdot \cdot 
\overline{R}^{A_{2nd-1}A_{2n}},  \label{Euler}
\end{equation}
which, by virtue of the Bianchi identity ($D\overline{R}^{ab}=0$), can be
shown to be closed, $d{\cal E}_{2n}=0$. [Here $\overline{R}^{AB}$ denotes
the ($d+1=2n$)-dimensional curvature, not to be confused with the $(d=2n-1)$%
-dimensional one, $R^{ab}$.] Substituting (\ref{Rbar}) into (\ref{Euler})
one finds the explicit form for ${\cal E}_{2n}$ in terms of the $(d=2n-1)$%
-dimensional tensors under the $SO(2n-2,1)$ (Lorentz) group, 
\[
{\cal E}_{2n}=n\epsilon
_{a_{1}...a_{2n-1}}(R^{a_{1}a_{2}}+l^{-2}e^{a_{1}}e^{a_{2}})\cdot \cdot
\cdot (R^{a_{2nd-3}a_{2n-2}}+l^{-2}e^{a_{2nd-3}}e^{a_{2n-2}})T^{a_{2n-1}}. 
\]

Now, the same arguments applied to find~the CS Lagrangian from the
Pontryagin density can be applied now to the Euler form. Thus, one may ask,
what is the $(2n-1)$-form whose exterior derivative gives (\ref{Euler})?.
Direct computation yields the answer in the form (\ref{Lovelock}), but with
fixed coefficients:

\begin{eqnarray}
{\cal E}_{2n} &=&d\left( n\sum\limits_{p=0}^{[d/2]}\frac{\kappa l^{2p-d}}{%
d-2p}\left( 
\begin{array}{l}
n-1 \\ 
p
\end{array}
\right) \epsilon _{a_{1}...a_{d}}R^{a_{1}a_{2}}\cdot \cdot \cdot
R^{a_{2p-1}a_{2p}}e^{a_{2p+1}}\cdot \cdot \cdot e^{a_{d}}\right) . \\
&=&dL_{2n-1}^{AdS}.
\end{eqnarray}
That is, the coefficients $\alpha _{p}$ given by (\ref{alpha2n-1}).

\subsection{Gauge invariance}

We now show that the Lagrangian $L_{2n-1}^{AdS}$ is really invariant under
infinitesimal $SO(d-1,2)$ (AdS) rotations. Under $SO(d-1,1)$ $e^{a}$ and $%
\omega _{\text{ }b}^{a}$ transform as 
\begin{equation}
\begin{array}{ll}
\omega ^{\text{ }ab}\rightarrow \omega ^{\prime \text{ }ab} & =d\lambda
^{ab}+\omega ^{ac}\lambda _{\text{ ~}c}^{b}+\omega ^{\text{ }cb}\lambda _{%
\text{ ~}c}^{a} \\ 
& \equiv D_{\omega }\lambda ^{ab} \\ 
e^{a}\rightarrow e^{\prime \text{ }a} & =\lambda _{\text{ ~}b}^{a}e^{b}
\end{array}
,  \label{Lortransf}
\end{equation}
then the new curvature $\overline{R}^{AB}$ transforms as a tensor under a
larger group. In fact, (\ref{Lortransf}) is a particular case of the $%
SO(d-1,2)$ transformations of the form 
\begin{equation}
\begin{array}{ll}
W^{AB}\rightarrow W^{^{\prime }\text{ }AB} & =d\Lambda ^{AB}+W_{\text{ }%
C}^{A}\Lambda ^{CB}+W_{\text{ }C}^{B}\Lambda ^{AC} \\ 
& \equiv D_{W}\Lambda ^{AB}
\end{array}
,  \label{AdStransf}
\end{equation}
where we have defined 
\begin{equation}
\Lambda ^{AB}=\left[ 
\begin{array}{ll}
\text{ }\lambda ^{ab} & l^{-1}\lambda ^{a} \\ 
-l^{-1}\lambda ^{b} & \text{ }0
\end{array}
\right] .  \label{Lambda}
\end{equation}
These transformations include, besides the Lorentz transformations
(determined by $\lambda ^{ab}$), also ``AdS boosts'' defined by $\lambda
^{a} $. Setting $\lambda ^{a}=0$ in (\ref{Lambda}), (\ref{Lortransf}) is
obtained. Therefore, the ($d+1$)-form ${\cal E}_{2n}$ is invariant under the
full Anti-de Sitter group. This in turn implies that the CS Lagrangian
obtained from the Euler form is also invariant under local AdS
transformations. It is the magic of the choice (\ref{alpha2n-1}) that the
group of local invariances of the action has grown from $SO(d-1,1)$ to $%
SO(d-1,2)$.

Finally, the Lagrangian in $d=2n-1$ dimensions reads

\begin{equation}
L_{2n-1}=\kappa \sum_{p=0}^{n-1}\frac{1}{D-2p}\left( 
\begin{array}{c}
n-1 \\ 
p
\end{array}
\right) l^{2p-D}\epsilon _{a_{1}\cdots a_{2n-1}}R^{a_{1}a_{2}}\cdots
R^{a_{2p-1}a_{2p}}e^{a_{2p+1}}\cdots e^{a_{2n-1}}.  \label{0}
\end{equation}

Here $\kappa $ is the gravitational constant similar to Newton's constant in 
$d=4$ and can be shown to be quantized in this theory \cite{Qkappa}. A
remarkable feature of this Lagrangian is that if written in terms of the AdS
connection $W^{AB}$, it has no dimensionful constants, and this makes it a
good candidate for a renormalizable field theory. However, there is an
obscure point here because the vacuum of the theory seems to be defined by
the {\bf AdS-flat }configuration $W^{AB}=0$ and this means that both $\omega
^{ab}$ and $e^{a}$ must vanish, a situation hard to reconcile with a
meaningful spacetime interpretation.

\subsection{Field Equations}

The field equations obtained varying with respect to $W^{AB}$ are 
\[
\epsilon _{A_{1}...A_{2n}}\overline{R}^{A_{3}A_{4}}\cdot \cdot \overline{R}%
^{A_{2nd-1}A_{2n}}=0, 
\]
or, equivalently varying with respect to $\omega ^{ab}$ and $e^{a}$%
\begin{equation}
\begin{array}{lll}
\delta \omega ^{ab}: & \epsilon
_{aba_{3}...a_{2n-1}}(R^{a_{3}a_{4}}+l^{-2}e^{a_{3}}e^{a_{4}})\cdot \cdot
\cdot (R^{a_{2nd-3}a_{2n-2}}+l^{-2}e^{a_{2nd-3}}e^{a_{2n-2}})T^{a_{2n-1}} & 
=0 \\ 
\delta e^{a}: & \epsilon
_{aa_{2}...a_{2n-1}}(R^{a_{2}a_{3}}+l^{-2}e^{a_{2}}e^{a_{3}})\cdot \cdot
\cdot (R^{a_{2nd-2}a_{2n-1}}+l^{-2}e^{a_{2nd-2}}e^{a_{2n-1}}) & =0
\end{array}
.  \label{Fieldeq}
\end{equation}

Configurations with~{\em vanishing local AdS curvature} ($\overline{R}%
^{AB}=0 $) are obvious solutions of these equations, while torsion-free
spaces with{\em \ flat Lorentz curvature} ($T^{a}=0=R^{ab}$) {\em are not}.
Other less trivial solutions are torsion-free spaces with ($2n-2$%
)-dimensional submanifolds of constant curvature foliated along one
direction, such as black holes or cosmological solutions \cite{JJG}.

\subsection{Gravity in 2+1 Dimensions}

As in all CS systems, the 2+1 dimensional case is illuminating and not
completely trivial. In this case the {\bf CS} Lagrangian whose exterior
derivative is the Euler density in 4 dimensions is the standard EH action in
three dimensions, 
\begin{equation}
L_{2+1}=\frac{\kappa }{l}\epsilon _{abc}(R^{ab}+\frac{1}{3l^{2}}%
e^{a}e^{b})e^{c}.  \label{L2+1}
\end{equation}
The corresponding field equations are 
\begin{eqnarray}
\epsilon _{abc}(R^{ab}+\frac{1}{l^{2}}e^{a}e^{b}) &=&0,  \label{2+1EE} \\
T^{a} &=&0.  \label{2+1T=0}
\end{eqnarray}

The last equation implies that the connection can be written as a function
of the vielbein and (\ref{2+1EE}) states that the spacetime must have
constant curvature at each point. In view of the fact that (2+1) gravity has
no propagating degrees of freedom, constant curvature spacetimes were
thought to be rather dull configurations. However, one can be surprised by
the fact that solving (\ref{2+1EE}) explicitly for spherically symmetric,
static configurations does not necessarily produce a globally AdS spacetime,
but an {\em AdS spacetime with identifications} as well. ~This is because
the field equations only refer to local properties of spacetime and do not
restrict the global topology further. Thus, the spacetime manifold~can be
cut and pasted, identifying points connected by a{\bf \ finite isometry}
along a Killing vector --as when one makes a cylinder out of a plane--, one
should still have a solution\footnote{%
Care should be taken, however not to produce closed timelike curves that
could generate paradoxical spacetimes scenarios (where one could kill his
ancestors, for example). Spaces with closed timelike curves of a {\em large
finite }proper length only could still make sense: if the period of those
curves is of the order of the age onf the universe, for instance.}. In fact,
one can produce a black hole in this fashion . What is even more remarkable,
is the fact that one can generate a solution by a Lorentz boost that sets
the black hole in rotation about its symmetry axis. \cite{MTZ}

A good example of a nontrivial solution of (\ref{2+1EE}) is the 2+1 black
hole geometry \cite{BTZ,BHTZ}, 
\begin{eqnarray}
ds^{2} &=&-N(r)^{2}f(r)^{2}dt^{2}+f(r)^{-2}dr^{2}+r^{2}(d\varphi +N^{\varphi
}(r)dt)^{2},  \nonumber \\
&&0\leq r<\infty ,\;\;\;\;0\leq \varphi <2\pi ,\;\;\;\;t_{1}\leq t\leq
t_{2}\,.
\end{eqnarray}

Solving the Einstein equations yields 
\begin{eqnarray}
f^{2} &=&r^{2}-M+\frac{J^{2}}{4r^{2}} \\
N &=&N(\infty ) \\
N^{\varphi } &=&-\frac{J}{2}\left( \frac{1}{r^{2}}+N^{\varphi }(\infty
)\right) \,,  \label{NphiJ}
\end{eqnarray}
which defines a black hole of mass $M$, angular momentum $J$, provided~$%
J^{2}\leq M^{2}$.

\subsection{\protect\medskip Exotic Gravity}

The Pontryagin form can be defined for any group, and in particular this is
also true of the Lorentz and AdS groups too. The Lorentz-Pontryagin form is
defined as 
\begin{equation}
P^{Lor}=R_{a_{2}}^{a_{1}}R_{a_{3}}^{a_{2}}\cdot \cdot \cdot
R_{a_{1}}^{a_{n}},  \label{PLor}
\end{equation}
and the AdS-Pontryagin form as 
\begin{equation}
P^{AdS}=\overline{R}_{A_{2}}^{A_{1}}\overline{R}_{A_{3}}^{A_{2}}\cdot \cdot
\cdot \overline{R}_{A_{1}}^{A_{n}}.  \label{PAdS}
\end{equation}

Because these curvature two-forms are antisymmetric in their indices, it is
obvious that they can only be defined for even $n$. This means that they are
naturally constructed in $4k$ dimensions. This in turn means that both AdS
and Lorentz Chern-Simons theories associated with the Pontryagin family can
only exist in $4k-\dot{1}$dimensions, these are the so-called {\bf exotic
Lagrangians}, which are independent from the Euler CS forms discussed above
in $d=2n-1$.

\section{Chern-Simons Supergravity}

Supersymmetry is the only nontrivial way to extend spacetime symmetries.
This result was well known before the era of supersymmetry and it states
that in a local field theory whose S-matrix relates in- and ~out-
eigenstates of energy and momentum, all other ``internal'' quantum numbers
such as color, flavor, hypercharge, etc., must be spacetime scalars. In
other words, under spacetime transformations these labels do not transform.
In mathematical terms this means that the group of invariances of the
S-matrix must be of the form $G={\cal S\otimes I}$, where ${\cal S}$ is the
group of spatial transformations (e.g., Poincar\'{e}, AdS, etc.), and ${\cal %
I}$ is the internal symmetry group.

For some time it was hoped that the nonrenormalizability of GR could be
cured by its supersymmetric extension. However, the initial hopes raised by
supergravity ({\bf SUGRA}) as a mechanism for taming the ultraviolet
divergences of pure gravity eventually vanished with the realization that
SUGRAs would be nonrenormalizable as well \cite{Townsend84}. Again, one can
see that, like GR, SUGRA is not a gauge theory for a group or a supergroup,
and that the local (super-) symmetry algebra closes naturally only on shell.
The algebra could be made to close off shell by force, at the cost of
introducing auxiliary fields --which are not guaranteed to exist for all $d$
and $N$ and \cite{RT}--, and still the theory would not have a fiber bundle
structure since the base manifold is identified with part of the fiber.

Whether it is the lack of fiber bundle structure the ultimate reason for the
nonrenormalizability of gravity remains to be proven. It is certainly true,
however, that if GR could be formulated as a gauge theory, the chances for
proving its renormalizability would clearly grow.

In three spacetime dimensions, on the other hand, both GR and SUGRA define
renormalizable quantum theories. It is strongly suggestive that precisely in
2+1 dimensions both theories can also be formulated as gauge theories on a
fiber bundle. It could be thought that the exact solvability miracle is due
to the absence of propagating degrees of freedom in three-dimensional
gravity, but final the power counting renormalizability argument rests on
the fiber bundle structure of the Chern-Simons form of those systems.

There are other known examples of gravitation theories in odd dimensions
which are genuine (off-shell) gauge theories for the anti-de Sitter ({\bf AdS%
}) or Poincar\'{e} groups \cite{Chamslett,Chamseddine,btz,Qkappa}. These
theories, as well as their supersymmetric extensions have propagating
degrees of freedom \cite{BGH} and are CS systems for the corresponding
groups as shown in \cite{btrz}.

\subsection{From Rigid Supersymmetry to Supergravity}

Rigid SUSY can be understood as an extension of the Poincar\'{e} algebra by
including supercharges which are the ``square roots'' of the generators of
rigid translations, $\{\bar{Q},Q\}\sim \Gamma \cdot \mbox{P}$. This idea can
be extended to local SUSY ~substituting the momentum P$_{\mu }=i\partial
_{\mu }$ by the generators of diffeomorphisms, ${\cal H}$, and relating them
to the supercharges by $\{\bar{Q},Q\}\sim \Gamma \cdot {\cal H}$. The
resulting theory has on-shell local supersymmetry algebra \cite{PvN}.

An alternative~approach --which we would like to advocate here-- is to
construct the supersymmetry {\em on the tangent space and not on the base
manifold}. This point of view is more natural if one recalls that spinors
provide a basis of irreducible representations for $SO(N)$, and not for $%
GL(N)$. Thus, spinors are naturally defined relative to a local frame on the
tangent space rather than on the coordinate basis. The basic point is to
reproduce the 2+1 ``miracle'' in higher dimensions. This idea has been
successfully applied in five dimensions\cite{Chamseddine}, and for pure
gravity \cite{btz,Qkappa} and to supergravity \cite{trz,btrz}. The SUGRA
construction has been carried out for spacetimes whose tangent space has AdS
symmetry \cite{trz}, and for its Poincar\'{e} contraction in \cite{btrz}.

\subsection{Assumptions of Standard Supergravity}

Three implicit assumptions are usually made in the construction of standard
SUGRA:

{\bf (i)} The fermionic and bosonic fields in the Lagrangian should come in
combinations such that their propagating degrees of freedom are equal in
number. This is usually achieved by adding to the graviton and the gravitini
a number of lower spin fields ($s<3/2$)\cite{PvN}. This matching, however,
is not necessarily true in AdS space, nor in Minkowski space if a different
representation of the Poincar\'{e} group (e.g., the adjoint representation)
is used \cite{Sohnius}.

The other two assumptions concern the purely gravitational sector. They are
as old as General Relativity itself and are dictated by economy: {\bf (ii)}
gravitons are described by the Hilbert action (plus a possible cosmological
constant), and, {\bf (iii)} the spin connection and the vielbein are not
independent fields but are related through the torsion equation. The fact
that the supergravity generators do not form a closed off-shell algebra can
be traced back to these assumptions..

The procedure behind {\bf (i)} is tightly linked to the idea that the fields
should be in a {\em vector} representation of the Poincar\'{e} group and
that the kinetic terms and couplings are such that the counting of degrees
of freedom works like in a minimally coupled gauge theory. This assumption
comes from the interpretation of supersymmetric states as represented by the
in- and out- plane waves in an asymptotically free, weakly interacting
theory in a minkowskian background. These conditions are not necessarily met
by a CS theory in an asymptotically AdS background. Apart from the
difference in background, which requires a careful treatment of the unitary
irreducible representations of the asymptotic symmetries \cite{Gunaydin},
the counting of degrees of freedom in CS theories is completely different
from the counting for the same connection 1-forms in a YM theory.

Thus, the only natural extension for a gauge theory of the AdS spacetime is
supergravity, constructed enlarging the group to some supergroups that
contain AdS but are otherwise semisimple. The construction of these theories
can be found in \cite{TrZ}. The crucial observation is that the Dirac
matrices provide a natural representation of the AdS algebra in any
dimension, thus, the connection $W^{AB}$ can be written in this
representation as $W=e^{a}J_{a}+\frac{1}{2}\omega ^{ab}J_{ab}$, where

\begin{equation}
J_{a}=\left[ 
\begin{array}{cc}
\frac{1}{2}(\Gamma _{a})_{\beta }^{\alpha } & 0 \\ 
0 & 0
\end{array}
\right] ,  \label{ja}
\end{equation}

\begin{equation}
J_{ab}=\left[ 
\begin{array}{cc}
\frac{1}{2}(\Gamma _{ab})_{\beta }^{\alpha } & 0 \\ 
0 & 0
\end{array}
\right] .  \label{jab}
\end{equation}

This spinorial representation for the connection naturally leads to a
representation for a superalgebra whose generators have entries in the
remaining blocks of similar matrices. The non zero blocks in (\ref{ja}) and (%
\ref{jab}) are $m\times m$ where $m=2^{[d/2]}$ is the number of components
of a spinor in $d$ dimensions. The algebra is completed by the (pseudo-)
Majorana generators of supersymmetry, 
\begin{equation}
Q_{\gamma }^{k}=\left[ 
\begin{array}{cc}
0 & \delta _{\gamma }^{\alpha }\delta _{j}^{k} \\ 
-C_{\gamma \beta }u^{ki} & 0
\end{array}
\right] ,  \label{Susygen}
\end{equation}
which are in a vector representation (described by the index $k$) of some
internal symmetry group with generators 
\begin{equation}
M^{kl}=\left[ 
\begin{array}{cc}
0 & 0 \\ 
0 & (m^{kl})_{j}^{i}
\end{array}
\right] .
\end{equation}
Following \cite{TrZ}, one arrives at all the possible superalgebras in
dimensions $d$, except for $d=5$ mod $\dot{4}$. In those cases the
representations are necessarily complex. Then the pseudo-Majorana condition
has to be relaxed and the generators of supersymmetry take the form 
\begin{eqnarray}
\bar{Q}_{\gamma }^{l} &=&\left[ 
\begin{array}{lr}
0 & \delta _{\gamma }^{\alpha }\delta _{j}^{l} \\ 
0 & 0
\end{array}
\right] ,  \label{Qbar} \\
Q_{\rho k} &=&\left[ 
\begin{array}{lr}
0 & 0 \\ 
-G_{\rho \beta }\delta _{k}^{i} & 0
\end{array}
\right] ,  \label{Qunbar}
\end{eqnarray}
instead of (\ref{Susygen}). With this, one can write the algebra for any $d$%
. The only nontrivial condition for the closure of the superalgebra comes
from the anticommutator of two supersymmetry generators. In some cases this
anticommutator includes all generators of the Pauli algebra, $\Gamma _{(k)}=%
\frac{1}{k!}\delta _{b_{1}\cdots b_{k}}^{a_{1}\cdots a_{k}}\Gamma
^{b_{1}}\cdots \Gamma ^{b_{k}},\;\;0\leq k\leq d$ (for $d=5$ mod $4$),
sometimes it contains only the symmetric subalgebra, $(C\Gamma
_{(k)})^{T}=C\Gamma _{(k)}$ (for $d=2,6,7,8$ mod $8)$, and sometimes it
includes only the antisymmetric ones, $(C\Gamma _{(k)})^{T}=-C\Gamma _{(k)}$
(for $d=2,3,4,6$ mod $8$). The fact that $d=2$ and $6$ appear in both
families is due to the fact that in those cases there are two inequivalent
choices of charge conjugation matrix ($C$) with $C^{T}=\pm C$.

We now consider the supersymmetric extensions of the locally AdS theories
defined above. In particular, one can write the ``exotic'' Lagrangian, 
\begin{equation}
d\widehat{L}_{4k-1}=\frac{-1}{2^{4k}}Tr[(R^{AB}\Gamma _{AB})^{2k}].
\label{LT}
\end{equation}
which is a particular form of the Pontryagin form (\ref{PAdS}). Other
possibilities of the form $Tr[\mbox{{\bf F}}^{n-p}]Tr[\mbox{{\bf F}}^{p}]$,
are not necessary to reproduce the minimal supersymmetric extensions of AdS
containing the Hilbert action. In the supergravity theories discussed below,
the gravitational sector is given by 
\[
dL_{2n-1}=STr[F 
\]
\begin{eqnarray}
dL_{2n-1}^{Grav} &=&\frac{-1}{2^{4k}}Tr[(R^{AB}\Gamma _{AB})^{2k}] \\
&=&\pm \frac{1}{2^{n}}L_{G\;2n-1}^{AdS}-\frac{1}{2}L_{T\;2n-1}^{AdS}.
\end{eqnarray}
which is a particular form of (\ref{PAdS}) where the trace over spinor
indices in this representation. Other possibilities of the form $%
\left\langle \mbox{{\bf F}}^{n-p}\right\rangle \left\langle \mbox{{\bf F}}%
^{p}\right\rangle $, are not necessary to reproduce the minimal
supersymmetric extensions of AdS containing the Hilbert action. The $\pm $
signs correspond to the two choices of inequivalent representations of $%
\Gamma $'s, which in turn reflect the two chiral representations in $d+1$.
As in the three-dimensional case, the supersymmetric extensions of $L_{2n-1}$
or any of the exotic Lagrangians such as $\widehat{L}_{2n-1}$, require using
both chiralities, thus doubling the algebras. Here we choose the + sign,
which gives the minimal superextension.

Under a gauge transformation, ${\bf A}$ transforms by $\delta {\bf A}=\nabla
\lambda $, where $\nabla $ is the covariant derivative for the same
connection ${\bf A}$. In particular, under a supersymmetry transformation, $%
\lambda =\bar{\epsilon}^{i}Q_{i}-\bar{Q}^{i}\epsilon _{i}$, and 
\begin{equation}
\delta _{\epsilon }{\bf A}=\left[ 
\begin{array}{cc}
\epsilon ^{k}\bar{\psi}_{k}-\psi ^{k}\bar{\epsilon}_{k} & D\epsilon _{j} \\ 
-D\bar{\epsilon}^{i} & \bar{\epsilon}^{i}\psi _{j}-\bar{\psi}^{i}\epsilon
_{j}
\end{array}
\right] ,  \label{delA}
\end{equation}
where $D$ is the covariant derivative on the bosonic connection, $D\epsilon
_{j}=(d+\frac{1}{2}[e^{a}\Gamma _{a}+\frac{1}{2}\omega ^{ab}\Gamma _{ab}+%
\frac{1}{r!}b^{[r]}\Gamma _{[r]}])\epsilon _{j}-a_{j}^{i}\epsilon _{i}$.

Two interesting cases can be mentioned here:

\subsection{d=5 SUGRA}

In this case the supergroup is $U(2,2|N)$. The associated connection can be
written as, 
\[
{\bf A=}e^{a}{\bf J}_{a}+\frac{1}{2}\omega ^{ab}{\bf J}_{ab}+a^{\Lambda }%
{\bf T}_{\Lambda }+(\bar{\psi}^{r}{\bf Q}_{r}-{\bf \bar{Q}}^{r}\psi _{r})+b%
{\bf Z}, 
\]
where the generators ${\bf J}_{a}$, ${\bf J}_{ab}$, form an AdS algebra ($%
so(4,2)$), ${\bf T}_{\Lambda }$ ($\Lambda =1,\cdot \cdot \cdot N^{2}-1$) are
the generators of $su(N)$, ${\bf Z}$ generates a $U(1)$ subgroup and ${\bf Q}%
,{\bf \bar{Q}}$ are the supersymmetry generators, which transform in a
vector representation of $SU(N)$. The Chern-Simons Lagrangian for this gauge
algebra is defined by the relation $dL=iSTr[{\bf F}^{3}]$, where ${\bf F=dA+}
$ ${\bf A}^{2}${\bf \ }is the (antihermitean) curvature. Using this
definition, one obtains the Lagrangian originally discussed by Chamseddine
in \cite{Chamseddine},

\begin{equation}
L=L_{G}(\omega ^{ab},e^{a})+L_{su(N)}(a_{s}^{r})+L_{u(1)}(\omega
^{ab},e^{a},b)+L_{F}(\omega ^{ab},e^{a},a_{s}^{r},b,\psi _{r}),  \label{L}
\end{equation}
with

\begin{equation}
\begin{array}{lll}
L_{G} & = & \frac{1}{8}\epsilon _{abcde}\left[ R^{ab}R^{cd}e^{e}/l+\frac{2}{3%
}R^{ab}e^{c}e^{d}e^{e}/l^{3}+\frac{1}{5}e^{a}e^{b}e^{c}e^{d}e^{e}/l^{5}%
\right] \\ 
L_{su(N)} & = & -Tr\left[ a(da)^{2}+\frac{3}{2}a^{3}da+\frac{3}{5}%
a^{5}\right] \\ 
L_{u(1)} & = & \left( \frac{1}{4^{2}}-\frac{1}{N^{2}}\right) b(db)^{2}+\frac{%
3}{4l^{2}}\left[ T^{a}T_{a}-R^{ab}e_{a}e_{b}-l^{2}R^{ab}R_{ab}/2\right] b \\ 
&  & +\frac{3}{N}f_{s}^{r}f_{r}^{s}b \\ 
L_{F} & = & 
\begin{array}{l}
\frac{3}{2i}\left[ \bar{\psi}^{r}{\cal R}\nabla \psi _{r}+\bar{\psi}^{s}%
{\cal F}_{s}^{r}\nabla \psi _{r}\right] +c.c.
\end{array}
\end{array}
,  \label{Li}
\end{equation}
where $a_{s}^{r}\equiv a^{\Lambda }(\tau _{\Lambda })_{s}^{r}$ is the $%
su(2,2)$ connection, $f_{s}^{r}$ is its curvature, and the bosonic blocks of
the supercurvature: ${\cal R}=\frac{1}{2}T^{a}\Gamma _{a}+\frac{1}{4}%
(R^{ab}+e^{a}e^{b})\Gamma _{ab}+\frac{i}{4}dbI-\frac{1}{2}\psi _{s}\bar{\psi}%
^{s}$, ${\cal F}_{s}^{r}=f_{s}^{r}+\frac{i}{N}db\delta _{s}^{r}-\frac{1}{2}%
\bar{\psi}^{r}\psi _{s}$. The cosmological constant is $-l^{-2},$ and the
AdS covariant derivative $\nabla $ acting on $\psi _{r}$ is

\begin{equation}
\nabla \psi _{r}=D\psi _{r}+\frac{1}{2l}e^{a}\Gamma _{a}\psi
_{r}-a_{\,r}^{s}\psi _{s}+i\left( \frac{1}{4}-\frac{1}{N}\right) b\psi _{r}.
\label{delta}
\end{equation}
where $D$ is the covariant derivative in the Lorentz connection.

The above relation implies that the fermions carry a $u(1)$ ``electric''
charge given by $e=\left( \frac{1}{4}-\frac{1}{N}\right) $. The purely
gravitational part, $L_{G}$ is equal to the standard Einstein-Hilbert action
with cosmological constant, plus the dimensionally continued Euler density%
\footnote{%
The first term in $L_{G}$ is the dimensional continuation of the Euler (or
Gauss-Bonnet) density from two and four dimensions, exactly as the
three-dimensional Einstein-Hilbert Lagrangian is the continuation of the the
two dimensional Euler density. This is the leading term in the limit of
vanishing cosmological constant ($l\rightarrow \infty )$, whose local
supersymmetric extension yields a nontrivial extension of the Poincar\'{e}
group \cite{btrz}.}.

The action is by construction invariant --up to a surface term-- under the
local (gauge generated) supersymmetry transformations $\delta _{\lambda
}A=-(d\lambda +[A,\lambda ])$ with $\lambda =\bar{\epsilon}^{r}{\bf Q}_{r}-%
{\bf \bar{Q}}^{r}\epsilon _{r}$, or

\[
\begin{array}{lll}
\delta e^{a} & = & \frac{1}{2}\left( \overline{\epsilon }^{r}\Gamma ^{a}\psi
_{r}-\bar{\psi}^{r}\Gamma ^{a}\epsilon _{r}\right) \\ 
\delta \omega ^{ab} & = & -\frac{1}{4}\left( \bar{\epsilon}^{r}\Gamma
^{ab}\psi _{r}-\bar{\psi}^{r}\Gamma ^{ab}\epsilon _{r}\right) \\ 
\delta a_{\,s}^{r} & = & -i\left( \bar{\epsilon}^{r}\psi _{s}-\bar{\psi}%
^{r}\epsilon _{s}\right) \\ 
\delta \psi _{r} & = & -\nabla \epsilon _{r} \\ 
\delta \bar{\psi}^{r} & = & -\nabla \bar{\epsilon}^{r} \\ 
\delta b & = & -i\left( \bar{\epsilon}^{r}\psi _{r}-\bar{\psi}^{r}\epsilon
_{r}\right) .
\end{array}
\]

As can be seen from (\ref{Li}) and (\ref{delta}), for $N=4$ the $b$ field
looses its kinetic term and decouples from the fermions (the gravitino
becomes uncharged with respect to the $U(1)$ field). The only remnant of the
interaction with the $b$ field is a dilaton-like coupling with the
Pontryagin four forms for the AdS and $SU(N)$ groups (in the bosonic
sector). As it is also shown in the Appendix A, the case $N=4$ is also
special at the level of the algebra, which becomes a superalgebra with a $%
u(1)$ central extension.

In the bosonic sector, for $N=4$, the field equation obtained from the
variation with respect to $b$ states that the Pontryagin four form of AdS
and $SU(N)$ groups are proportional . Consequently, if the curvatures
approach zero sufficiently fast at spatial infinity, there is a conserved
topological current which states that, for the spatial section, the second
Chern characters of AdS\ and $SU(4)$ are proportional. Consequently, if the
spatial section has no boundary, the corresponding Chern numbers are
related. Using the fact that $\Pi _{4}(SU(4))=0$, the above implies that the
Hirzebruch signature plus the Nieh-Yan number of the spatial section cannot
change in time.

\subsection{d=11 SUGRA}

In this case, the smallest AdS superalgebra is $osp(32|1)$ and the
connection is {\bf A} =$\frac{1}{2}\omega ^{ab}J_{ab}+e^{a}J_{a}+\frac{1}{5!}%
b^{abcde}J_{abcde}+\bar{Q}\psi $, where $b$ is a totally antisymmetric
fifth-rank Lorentz tensor one-form. Now, in terms of the elementary bosonic
and fermionic fields, the CS form in (\ref{Fn}) reads 
\begin{equation}
L_{11}^{osp(32|1)}({\bf A})=L_{11}^{sp(32)}(\Omega )+L_{F}(\Omega ,\psi ),
\label{L11}
\end{equation}
where $\Omega \equiv \frac{1}{2}(e^{a}\Gamma _{a}+\frac{1}{2}\omega
^{ab}\Gamma _{ab}+\frac{1}{5!}b^{abcde}\Gamma _{abcde})$ is an $sp(32)$
connection. The bosonic part of (\ref{L11}) can be written as 
\[
L_{11}^{sp(32)}(\Omega )=2^{-6}L_{G\;11}^{AdS}(\omega ,e)-\frac{1}{2}%
L_{T\;11}^{AdS}(\omega ,e)+L_{11}^{b}(b,\omega ,e). 
\]
The fermionic Lagrangian is 
\begin{eqnarray*}
L_{F} &=&6(\bar{\psi}R^{4}D\psi )-3\left[ (D\bar{\psi}D\psi )+(\bar{\psi}%
R\psi )\right] (\bar{\psi}R^{2}D\psi ) \\
&&-3\left[ (\bar{\psi}R^{3}\psi )+(D\bar{\psi}R^{2}D\psi )\right] (\bar{\psi}%
D\psi )+ \\
&&2\left[ (D\bar{\psi}D\psi )^{2}+(\bar{\psi}R\psi )^{2}+(\bar{\psi}R\psi )(D%
\bar{\psi}D\psi )\right] (\bar{\psi}D\psi ),
\end{eqnarray*}
where $R=d\Omega +\Omega ^{2}$ is the $sp(32)$ curvature. The supersymmetry
transformations (\ref{delA}) read 
\[
\begin{array}{lll}
\delta e^{a}=\frac{1}{8}\bar{\epsilon}\Gamma ^{a}\psi & \hspace{1cm} & 
\delta \omega ^{ab}=-\frac{1}{8}\bar{\epsilon}\Gamma ^{ab}\psi \\ 
&  &  \\ 
\delta \psi =D\epsilon & \hspace{1cm} & \delta b^{abcde}=\frac{1}{8}\bar{%
\epsilon}\Gamma ^{abcde}\psi .
\end{array}
\]

Standard eleven-dimensional supergravity \cite{CJS} is an N=1 supersymmetric
extension of Einstein-Hilbert gravity that cannot accommodate a cosmological
constant \cite{BDHS,Deser}. An $N>1$ extension of this theory is not known.
In our case, the cosmological constant is necessarily nonzero by
construction and the extension simply requires including an internal $so(N)$
gauge field coupled to the fermions, and the resulting Lagrangian is an $%
osp(32|N)$ CS form \cite{tronco}.

\subsection{Summary}

The supergravities presented here have two distinctive features: The
fundamental field is always the connection {\bf A} and, in their simplest
form, they are pure CS systems (matter couplings are discussed below). As a
result, these theories possess a larger gravitational sector, including
propagating spin connection. Contrary to what one could expect, the
geometrical interpretation is quite clear, the field structure is simple
and, in contrast with the standard cases, the supersymmetry transformations
close off shell without auxiliary fields.

{\bf Torsion.} It can be observed that the torsion Lagrangians ($L_{T}$)are
odd while the torsion-free terms ($L_{G}$) are even under spacetime
reflections. The minimal supersymmetric extension of the AdS group in $4k-1$
dimensions requires using chiral spinors of $SO(4k)$ \cite{Gunaydin}. This
in turn implies that the gravitational action has no definite parity, but
requires the combination of $L_{T}$ and $L_{G}$ as described above. In $%
D=4k+1$ this issue doesn't arise due to the vanishing of the torsion
invariants, allowing constructing a supergravity theory based on $L_{G}$
only, as in \cite{Chamseddine}. If one tries to exclude torsion terms in $%
4k-1$ dimensions, one is forced to allow both chiralities for $SO(4k)$
duplicating the field content, and the resulting theory has two copies of
the same system \cite{Horava}.

{\bf Field content and extensions with N$>$1.} The field content compares
with that of the standard supergravities in $D=5,7,11$ as follows:\newline

\begin{center}
\begin{tabular}{c|c|l|l|}
\cline{2-4}
\hspace{.7cm} & D & Standard supergravity & CS supergravity \\ \cline{2-4}
& 5 & $e^a_{\mu}$ $\psi^{\alpha}_{\mu}$ $\bar{\psi}_{\alpha \mu}$ & $%
e_{\mu}^a$ $\omega^{ab}_{\mu}$ $\psi^{\alpha}_{\mu}$ $\bar{\psi}_{\alpha
\mu} $ $b$ \\ \cline{2-4}
& 7 & $e^a_{\mu}$ $A_{[3]}$ $\psi^{\alpha i}_{\mu}$ $a^i_{\mu j}$ $%
\lambda^{\alpha}$ $\phi$ & $e_{\mu }^a$ $\omega ^{ab}_{\mu}$ $%
\psi_{\mu}^{\alpha i }$ $a_{\mu j}^i$ \\ \cline{2-4}
& 11 & $e^a_{\mu}$ $A_{[3]}$ $\psi^{\alpha}_{\mu}$ & $e_{\mu}^{a}$ $\omega
^{ab}_{\mu }$ $\psi_{\mu }^{\alpha }$ $b^{abcde}_{\mu }$ \\ \cline{2-4}
\end{tabular}
\end{center}

Standard supergravity in five dimensions is dramatically different from the
theory presented here, which was also discussed by Chamseddine in \cite
{Chamseddine}.

Standard seven-dimensional supergravity is an $N=2$ theory (its maximal
extension is N=4), whose gravitational sector is given by Einstein-Hilbert
gravity with cosmological constant and with a background invariant under $%
OSp(2|8)$ \cite{D=7,Salam-Sezgin}. Standard eleven-dimensional supergravity 
\cite{CJS} is an N=1 supersymmetric extension of Einstein-Hilbert gravity
that cannot accommodate a cosmological constant \cite{BDHS,Deser}. An $N>1$
extension of this theory is not known.

In the case presented here, the extensions to larger $N$ are straightforward
in any dimension. In $D=7$, the index $i$ is allowed to run from $2$ to $2s$%
, and the Lagrangian is a CS form for $osp(2s|8)$. In $D=11$, one must
include an internal $so(N)$ field and the Lagrangian is an $osp(32|N)$ CS
form \cite{trz}. The cosmological constant is necessarily nonzero in all
cases.

{\bf Spectrum.} The stability and positivity of the energy for the solutions
of these theories is a highly nontrivial problem. As shown in Ref. \cite{BGH}%
, the number of degrees of freedom of bosonic CS systems for $D\geq 5$ is
not constant throughout phase space and different regions can have radically
different dynamical content. However, in a region where the rank of the
symplectic form is maximal the theory behaves as a normal gauge system, and
this condition is stable under perturbations. As it is shown in \cite{CTZ},
there exists a nontrivial extension of the AdS superalgebra with one abelian
generator for which anti-de Sitter space without matter fields is a
background of maximal rank, and the gauge superalgebra is realized in the
Dirac brackets. For example, for $D=11$ and $N=32$, the only nonvanishing
anticommutator reads 
\begin{eqnarray*}
\{Q_{\alpha }^{i},\bar{Q}_{\beta }^{j}\} &=&\frac{1}{8}\delta ^{ij}\left[
C\Gamma ^{a}J_{a}+C\Gamma ^{ab}J_{ab}+C\Gamma ^{abcde}Z_{abcde}\right]
_{\alpha \beta } \\
&&-M^{ij}C_{\alpha \beta },
\end{eqnarray*}
where $M^{ij}$ are the generators of $SO(32)$ internal group. On this
background the $D=11$ theory has $2^{12}$ fermionic and $2^{12}-1$ bosonic
degrees of freedom. The (super)charges obey the same algebra with a central
extension. This fact ensures a lower bound for the mass as a function of the
other bosonic charges \cite{GH}.

{\bf Classical solutions.} The field equations for these theories in terms
of the Lorentz components ($\omega $, $e$, $b$, $a$, $\psi $) are spread-out
expressions for $<${\bf F}$^{n-1}G_{(a)}>=0$, where $G_{(a)}$ are the
generators of the superalgebra. It is rather easy to verify that in all
these theories the anti-de Sitter space is a classical solution , and that
for $\psi =b=a=0$ there exist spherically symmetric, asymptotically AdS
standard \cite{JJG}, as well as topological \cite{ABHPB} black holes. In the
extreme case these black holes can be shown to be BPS states.

{\bf Matter couplings.} It is possible to introduce a minimal couplings to
matter of the form {\bf A}$\cdot ${\bf J}. For $D=11$, the matter content is
that of a theory with (super-) 0, 2, and 5--branes, whose respective
worldhistories couple to the spin connection and the $b$ fields.

{\bf Standard SUGRA.} Some sector of these theories might be related to the
standard supergravities if one identifies the totally antisymmetric part of
the contorsion tensor in a coordinate basis, $k_{\mu \nu \lambda }$, with
the abelian 3-form, $A_{[3]}$. In 11 dimensions one could also identify the
antisymmetric part of $b$ with an abelian 6-form $A_{[6]}$, whose exterior
derivative, $dA_{[6]}$, is the dual of $F_{[4]}=dA_{[3]}$. Hence, in $D=11$
the CS theory possibly contains the standard supergravity as well as some
kind of dual version of it.

\section{Dynamical Contents of Chern Simons Theories}

The physical meaning of a theory is defined by the dynamics it displays both
at the classical as well as at the quantum levels. In order to understand
this question one should be able to separate the physical degrees of freedom
from those which are redundant. In particular, it should be possible --at
least in principle-- to separate the propagating modes from the gauge
degrees of freedom, and from those which do not evolve independently at all
(second class constraints). The standard way to proceed is Dirac's
constrained Hamiltonian analysis and has been studied in {\bf CS} systems in 
\cite{BGH}. Here we summarize this analysis but refer the reader to the
original papers for the details.

\subsection{The BGH construction}

From the dynamical point of view, a CS system can be described by a
Lagrangian of the form \footnote{%
Note that in this section, for notational simplicity, we assume the
spacetime to be ($2n+1$)-dimensional.} 
\begin{equation}
L_{2n+1}=l_{a}^{i}(A_{j}^{b})\dot{A}_{i}^{a}-A_{o}^{a}K_{a},
\end{equation}
where the ($2n+1$)-dimensional spacetime has been split into space and time,
and 
\[
K_{a}=-\frac{1}{2^{n}n}\gamma _{aa_{1}....a_{n}}\epsilon
^{i_{1}...i_{2n}}F_{i_{1}i_{2}}^{a_{1}}\cdot \cdot \cdot
F_{i_{2n-1}i_{2n}}^{a_{n}}. 
\]
The field equations are 
\begin{eqnarray}
\Omega _{ab}^{ij}(\dot{A}_{j}^{b}-D_{j}A_{0}^{b}) &=&0,  \label{csEq} \\
K_{a} &=&0,  \label{K=0}
\end{eqnarray}
where 
\begin{eqnarray}
\Omega _{ab}^{ij} &=&\frac{\delta l_{b}^{j}}{\delta A_{i}^{a}}-\frac{\delta
l_{a}^{i}}{\delta A_{j}^{b}}  \label{symplectic} \\
&=&-\frac{1}{2^{n-1}}\gamma _{aba_{2}....a_{n}}\epsilon
^{iji_{3}...i_{2n}}F_{i_{3}i_{4}}^{a_{2}}\cdot \cdot \cdot
F_{i_{2n-1}i_{2n}}^{a_{n}}  \nonumber
\end{eqnarray}
is the symplectic matrix. The passage to the Hamiltonian has the problem
that the velocities appear linearly in the Lagrangian, ant therefore there
is a number of primary constraints 
\begin{equation}
\phi _{a}^{i}\equiv p_{a}^{i}-l_{a}^{i}\approx 0.  \label{Phi's}
\end{equation}

Besides these there are the secondary constraints $K_{a}\approx 0$, which
can be combined with the $\phi $'s into the expressions 
\begin{equation}
G_{a}\equiv -K_{a}+D_{i}\phi _{a}^{i},  \label{Ga}
\end{equation}
which together with the $\phi $'s form a closed algebra, 
\[
\begin{array}{ll}
\{\phi _{a}^{i},\phi _{b}^{j}\} & =\Omega _{ab}^{ij} \\ 
\{\phi _{a}^{i},G_{b}\} & =C_{ab}^{c}\phi _{c}^{i} \\ 
\{G_{a},G_{b}\} & =C_{ab}^{c}G_{c}
\end{array}
, 
\]
where $C_{ab}^{c}$ are the structure constants of the gauge algebra of the
theory. Clearly the $G$'s form a first class algebra which reflects the
gauge invariance of the theory, while some of the $\phi $'s are second class
and some are first class, depending on the rank of the symplectic form. Here
we face the first serious difficulty with CS theory: the matrix $\Omega
_{ab}^{ij}$depends on the field configurations, and therefore its rank
cannot be thought of as a constant, but it can change from one region of
phase space to another. This issue has been analyzed in the context of some
simplified mechanical models and the conclusion is that the degeneracy of
the system occurs at surfaces of lower dimensionality in phase space, which
can be viewed as end sets of (unstable) initial points or sets of (stable)
end points for the evolution. Unless the system is chaotic, it can be
expected that generic configurations where the rank of $\Omega _{ab}^{ij}$
is maximal should fill most of phase space \cite{STZ}.

There is a second more tractable problem and that is how to separate the
first and second class constraints among the $\phi $'s. In Ref.\cite{BGH}
the following results are shown:

\begin{itemize}
\item  The maximal rank of $\Omega _{ab}^{ij}$ is $2nN-2n$ , where $N$ is
the number of generators in the gauge Lie algebra.

\item  In consequence, there are $2n$ first class constraints among the $%
\phi $'s which correspond to the generators of spatial diffeomorphisms ($%
{\cal H}_{i}$).

\item  The generator of timelike reparametrizations ${\cal H}_{\perp }$ is
not an independent first class constraint.
\end{itemize}

Putting all these facts together, one concludes that in a generic
configuration, the number of degrees of freedom of the theory is 
\begin{eqnarray}
g &=&(\text{n}{{}^\circ}\text{ of coordinates})-(\text{n}{{}^\circ}\text{ of
1}^{st}\text{class constraints})-\frac{1}{2}(\text{n}{{}^\circ}\text{ of
second class constraints})  \nonumber \\
&=&2nN-(N+2n)-\frac{1}{2}(2nN-2n)=nN-N-n.  \label{nN-N-n}
\end{eqnarray}

This result is somewhat perplexing. For the gravity theory as a CS system
for the AdS group, this result gives, in $d=2n+1$ dimensions has 
\[
N=\frac{(2n+1)(2n+2)}{2}=(2n+1)(n+1), 
\]
and therefore, 
\begin{eqnarray}
g_{AdS}^{CS} &=&n(2n+1)(n+1)-(2n+1)(n+1)-n  \nonumber \\
&=&2n^{3}+n^{2}-3n-1.  \label{n3}
\end{eqnarray}
This number of propagating degrees of freedom is much larger than the number
found in a purely metric theory of gravity in dimension $d$. 
\begin{eqnarray}
g_{d}^{metric} &=&\frac{d(d-3)}{2}=\frac{(2n+1)(2n-2)}{2}  \nonumber \\
&=&2n^{2}-n-1.  \label{n2}
\end{eqnarray}

Thus, for $d\geq 3$ the number of degrees of freedom of the CS theory is
much larger than that of the standard metric theory. In particular, for $d=5$%
, (\ref{n3}) is 13, while (\ref{n2}) equals 5. Obviously, the extra degrees
of freedom must correspond to propagating modes contained in the torsion,
which here are independent from the metric degrees of freedom. The precise
identification of the propagating degrees freedom, however requires
separating the first and second class constraints, inverting the Dirac
matrix and eliminating the second class constraints consistently. This
objective seems unattainable in general for an arbitrary gauge group.

As it is also shown in \cite{BGH}, an important simplification occurs when
the group has an invariant abelian factor. In that case the symplectic
matrix $\Omega _{ab}^{ij}$ takes a partially block-diagonal form where the
kernel has the maximal size allowed by a generic configuration. For example,
in five dimensions, those authors show that for a group $G=G_{0}\otimes U(1)$%
, one has a generic configuration if the curvature for the $G_{0}$
connection vanish, but the $U(1)$ curvature, $f=db$, is a nondegenerate
2-form,

. 
\[
\left. \Omega _{ab}^{ij}\right| _{F^{a}=0}=\left( 
\begin{array}{ll}
0 & 0 \\ 
0 & -\frac{1}{2}\gamma _{ab}\epsilon ^{ijkl}f_{kl}
\end{array}
\right) . 
\]
Then, the lower block of $\Omega _{ab}^{ij}$ is the Dirac matrix and the
constrained Hamiltonian analysis can proceed in the standard way.

It is a nice surprise in the cases of CS supergravities discussed above that
it seems that for certain unique choices extended supesymmetries the
algebras develop an abelian subalgebra and make the separation of first and
second class constraints possible. It is remarkable that in some cases
(e.g., for $d=5$, $N=4$) the algebra {\bf is not} a direct sum but an
algebra with an abelian central extension. In other cases (e.g., for $d=11$, 
$N=32$), the algebra is a direct sum, but the abelian subgroup is not put in
by hand but it is a subset of the generators that for that particular
extended supersymmetry spontaneously decouples from the rest of the algebra.

\subsection{Final Comments}

Chern-Simons theories contain a wealth of other interesting features,
starting with its relation to geometry and field theories and knot
invariants. The higher-dimensional CS systems remain somewhat mysterious
especially because of the difficulties to treat them as quantum theories.
However, they have many ingredients that make them likely models to be
quantized: They carry no dimensionful couplings; the only parameters they
can have must be quantized; in gravity these are the only theories of the
Lovelock family that gives rise to black holes with positive specific heat 
\cite{BHscan} and hence, capable to reach thermal equilibrium with an
external heat bath.

Efforts to quantize CS systems seem promising at least in the cases in which
the space admits a complex structure so that the symplectic form can be cast
as a K\"{a}hler form \cite{NS}

{\bf ACKNOWLEDGMENTS}

The author is grateful to R.Aros, M.Henneaux,C.Mart\'{i}nez, F.M\'{e}ndez,
R.Olea, C.Teitelboim and R.Troncoso, for many enlightening discussions and
helpful comments. This work was supported in part by grants 1980788 and
1990189 from FONDECYT (Chile), and 27.953/ZI-DICYT (USACH). Institutional
support to CECS from Fuerza A\'{e}rea de Chile, I.Municipalidad de Las
Condes, I. Municipalidad de Santiago, and a group of Chilean companies (AFP
Provida, Business Design Associates, CGE, Codelco, Copec, Empresas CMPC,
Gener SA, Minera Collahuasi, Minera Escondida, Novagas and Xerox -Chile) is
also recognized. CECS is a Millennium Science Institute.

\renewcommand{\theequation}{A.\arabic{equation}} \setcounter{equation}{0}

\end{document}